\documentclass[lettersize,journal]{IEEEtran}
\usepackage{amsmath,amsfonts}
\usepackage{algorithmic}
\usepackage{array}
\usepackage{textcomp}
\usepackage{stfloats}
\usepackage[colorlinks=false, pdfborder={0 0 0}]{hyperref}
\usepackage{verbatim}
\usepackage{graphicx}
\usepackage{cite}
\usepackage{gensymb}
\usepackage{subfigure}
\def\BibTeX{{\rm B\kern-.05em{\sc i\kern-.025em b}\kern-.08em
    T\kern-.1667em\lower.7ex\hbox{E}\kern-.125emX}}
\usepackage{balance}
\usepackage[dvipsnames]{xcolor}

\definecolor{bluerevision}{RGB}{0,112,192}

\makeatletter
 \let\old@ps@headings\ps@headings
 \let\old@ps@IEEEtitlepagestyle\ps@IEEEtitlepagestyle
 \def\confheader#1{%
 \def\ps@headings{%
 \old@ps@headings%
 \def\@oddhead{\strut\hfill#1\hfill\strut}%
 \def\@evenhead{\strut\hfill#1\hfill\strut}%
 }%
 \def\ps@IEEEtitlepagestyle{%
 \old@ps@IEEEtitlepagestyle%
 \def\@oddhead{\strut\hfill#1\hfill\strut}%
 \def\@evenhead{\strut\hfill#1\hfill\strut}%
 }%
 \ps@headings%
 }
 \makeatother
\confheader{%
This work has been accepted for publication in \textit{IEEE Wireless Communications Letters}. DOI: 10.1109/LWC.2024.3434415
}

 \usepackage[pscoord]{eso-pic}
\newcommand{\placetextbox}[3]{
 \setbox0=\hbox{#3}
 \AddToShipoutPictureFG*{ \put(\LenToUnit{#1\paperwidth},\LenToUnit{#2\paperheight}){\vtop{{\null}\makebox[0pt][c]{#3}}}
 }
 }
 \placetextbox{.5}{0.055}{\small{\copyright 2024 IEEE. Personal use of this material is permitted. Permission from IEEE must be obtained for all other uses, in any current or future media,}}
 \placetextbox{.5}{0.045}{\small{including reprinting/republishing this material for advertising or promotional purposes, creating new collective works, for resale or redistribution to}}
 \placetextbox{.5}{0.035}{\small{servers or lists, or reuse of any copyrighted component of this work in other works.}}

\begin{document}

\title{Observations on Large-Scale Attenuation Effects \\ in a 26 GHz Urban Micro-Cell Environment}

\author{Alejandro Ramírez-Arroyo, Troels B. Sørensen, Peter Beltoft, Henrik Christiansen,\\ Juan F. Valenzuela-Valdés, Preben Mogensen 

\thanks{Alejandro Ramírez-Arroyo, Troels B. Sørensen and Preben Mogensen are with the Department of Electronic Systems, Aalborg University (AAU), 9220 Aalborg, Denmark (e-mail: araar@es.aau.dk; tbs@es.aau.dk; pm@es.aau.dk).}

\thanks{Peter Beltoft and Henrik Christiansen  are with Department of Radio Access Networks, TDC Net, 0900 Copenhagen, Denmark (e-mail: pbel@tdcnet.dk; henrch@tdcnet.dk)}

\thanks{Juan F. Valenzuela-Valdés is with the Research Centre for Information and Communication Technologies, Department of Signal Theory, Telematics and Communications, Universidad de Granada (UGR), 18071 Granada, Spain (e-mail: juanvalenzuela@ugr.es).}}

\markboth{Ramírez-Arroyo \MakeLowercase{\textit{et al.}}: Observations on Large-Scale Attenuation Effects in a 26 GHz Urban Micro-Cell Environment}%
{Ramírez-Arroyo \MakeLowercase{\textit{et al.}}: Observations on Large-Scale Attenuation Effects in a 26 GHz Urban Micro-Cell Environment}

\maketitle

\begin{abstract}
This letter presents a measurement campaign carried out in an FR2 urban outdoor environment in a live experimental network deployment. The radio propagation analysis from a physical perspective at 26 GHz is essential for the correct deployment and dimensioning of future communication networks. This study performs a walk test emulating realistic conditions under which a pedestrian may be affected, summarizing and evaluating some of the typical effects encountered in a communications scenario such as penetration losses in a building, losses due to vegetation or the human body, or diffraction/scattering propagation around corners in street canyon-like environments. The operational conditions of the 5G network, the urban micro-cell scenario, and the use of omnidirectional antennas on the UE side validate the channel conditions from a perspective closer to a realistic scenario for a pedestrian within a FR2 live network.
\end{abstract}


\section{Introduction}
\IEEEPARstart{W}{ireless} communications have undergone exponential growth over the last decades. This is mainly due to the objective of enhancing mobile services, such as the throughput in the network. Within the 5G generation mobile network, one promising opportunity lies in the millimeter wave (mmWave) paradigm. This range of the electromagnetic spectrum is less saturated than the sub-6 GHz spectrum due to the spectrum availability, enabling the allocation of several services with large bandwidths. Nevertheless, the use of high frequencies implies some challenges, such as higher propagation losses or paradigm shifts in propagation mechanisms such as wave diffraction or scattering.

For the aforementioned reasons, it is crucial to conduct radio propagation characterization and modeling at these frequencies. Multiple efforts have already been undertaken through the 3rd Generation Partnership Project (3GPP) and the International Telecommunication Union (ITU) with the frequency range 2 (FR2) band standardization, which corresponds to the bandwidth between 24.25~GHz to 71.0~GHz \cite{m2150, ts38211, ts38213}. Over the years, modeling work has been carried out from both theoretical perspectives (Saleh-Valenzuela, WINNER II, 3GPP SCM, Zwick models) and empirical approaches (Okumura-Hata, COST 231 models) \cite{channel_modeling_theoretical}. One of the main environments to be characterized are urban scenarios. Specifically, mobile networks are most in demand in densely populated urban outdoor environments.

From the experimental perspective, the research community has focused on the mmWave band characterization because of its potential for future networks \cite{channel_modeling_empirical}. These studies show how challenging urban scenarios are, where street furniture, as well as buildings, tend to constantly obstruct the signal received by the end user. For instance, large-scale modeling of densely urban scenarios like Manhattan has been performed \cite{5G_it_will_work}.  Subsequently, the focus has shifted to more specific phenomena such as propagation around corners \cite{diffraction_scattering_3}, penetration losses in buildings for Outdoor-to-Indoor (O2I) communications \cite{glass_2}, or the vegetation impact on the received power \cite{foliage_2}. Because the research is being carried out at the early stages of 5G technology, the studies are based on networks assembled with dedicated equipment specific to the analysis. Currently, as network deployments are starting to take place, the first studies based on live commercial networks are appearing, although mainly focused in the sub-6GHz band \cite{operational_sub6}.

In this letter, the analysis of a measurement campaign is performed in a 5G FR2 network in a live experimental network deployment. The study aims to show some of the typical large-scale propagation phenomena that an end-user may be exposed to in an urban micro-cell environment, and to which extent these effects are or can be represented by commonly accepted models. More specifically, the work focuses on the analysis of the path gain and the power received by and end-user throughout conditions such as Non Line-of-Sight propagation, vegetation blockage, around-corner propagation and building penetration. This research, which is based on a walk test, emulates the movements of a pedestrian in an urban scenario. Thus, the analysis of the measurements conducted in a live network with an omnidirectional antenna on the user equipment side, allows us to assess the effects a pedestrian may experience in a realistic urban scenario.

\section{Measurement Equipment and Scenario Description}

\noindent This work analyzes the cellular deployment in an FR2 urban outdoor environment. Particularly, the study is carried out in Sydhavnen, a residential district in Copenhagen, Denmark. 

Regarding the measurement setup, a base station located at a 34.9 m height operates according to 3GPP standards for 5G NR \cite{ts38211, ts38213, ts381014}. On the transmitter/base station (TX/BS) side, the smallest unit of information is provided by a single subcarrier and an OFDM symbol, which define a resource element (RE) given the frequency-time domain. These RE are grouped into 14 symbols and 12 subcarriers in order to form the physical resource blocks (PRBs). In 5G NR, flexibility is allowed according to the subcarrier spacing (SCS), which varies from 15~kHz (case A) to 240~kHz (case E). In this study, this value is set to 120 kHz, which corresponds to case D with NR numerology $\mu = 3$. The base station transmits over $B$~$=$~$100$~MHz bandwidth with an operation frequency $f = 26$~GHz corresponding to the n258 band in FR2. Specifically, the TX is a dual-polarized mmWave 5G NR Antenna Integrated Radio Ericsson AIR 5322 B258. This phased array antenna module, with $\pm60\degree$ and $\pm15\degree$ scan range in horizontal and vertical planes respectively, is arranged so that it is expected to cover the main avenue of the district as well as the surrounding areas. For additional information about the transmitter, the reader is referred to~\cite{TX_antenna}. Fig. 1 shows the location of the BS, as well as the residential area where coverage is expected to be provided. On the receiver/user equipment (RX/UE) side, a R\&S TSME6 radio test scanner is connected to an H-plane omnidirectional antenna (HPBW$_{E_{plane}} = 38\degree$) operating in the range between 3 GHz and 40 GHz. With a periodicity of 20 ms, the BS transmits a pilot signal known as Synchronization Signal Block (SSB). Within this signal is the Secondary Synchronization Signal (SSS), transmitted over 127 subcarriers. Thus, the radio scanner acquires the Reference Signal Received Power (RSRP) as the linear average of the RE received power over the SSS bandwidth, i.e. 15.24 MHz for SCS $=$ 120 kHz. The RX is kept at a height of 1.7 m above the ground. Additionally, the radio scanner is equipped with a GPS antenna, which allows tracking of the position where the measurements have been acquired. The measurements are based on a walk test with a pedestrian walking at an average speed of 1.5 m/s (5.4 km/h). Considering the 20 ms periodicity of the SSB, an average of 33.3 samples/m are captured.

\begin{figure}[t]
	\centering
	\includegraphics[width= 1\columnwidth]{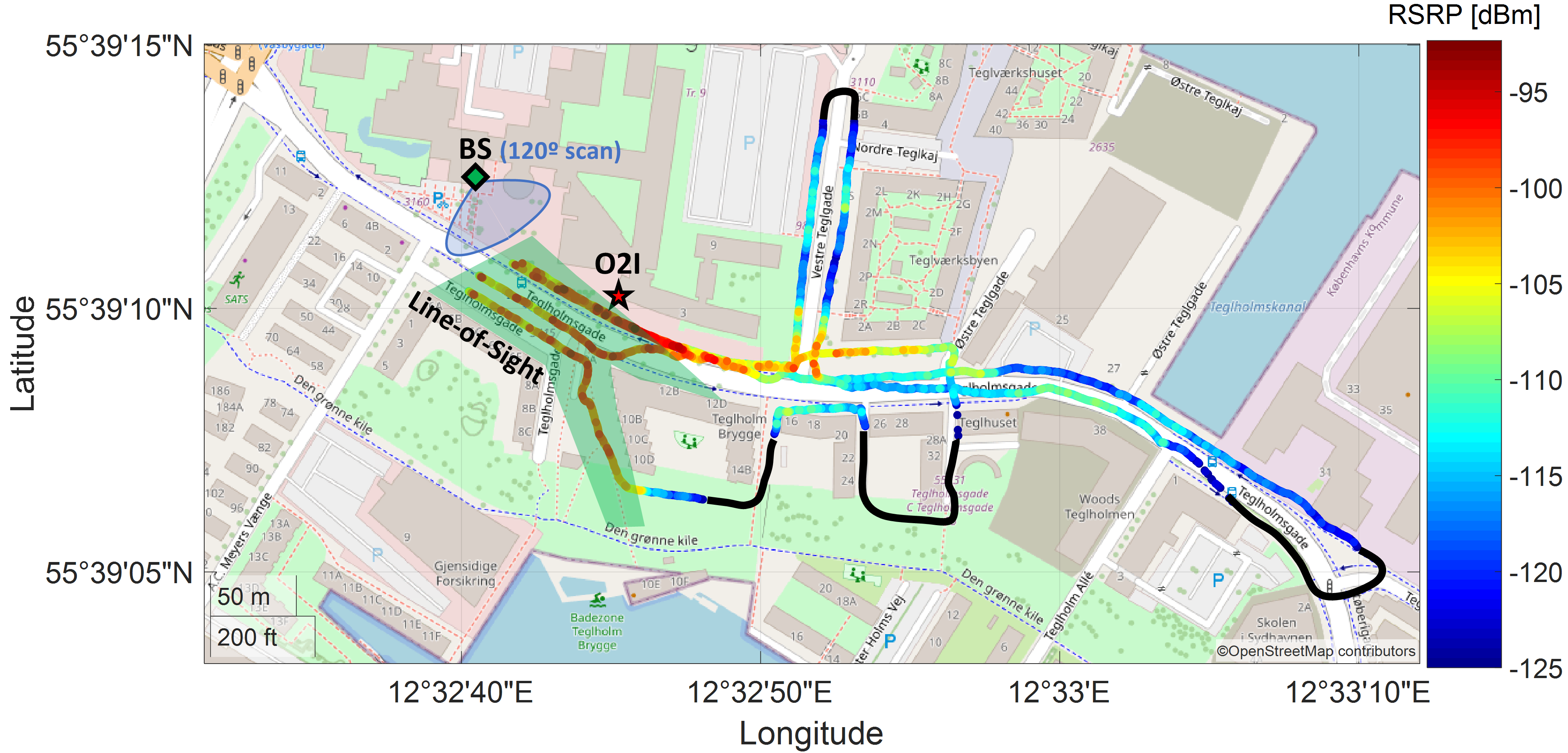}
	\caption{RSRP throughout the analyzed deployment. BS location, and its horizontal scan range, is pointed out with a green rhombus, while the Line-of-Sight region is marked within a green area. Outdoor-to-Indoor (O2I) analysis is performed in the red star position.} 
	\label{fig1}
\end{figure}

Regarding the environment, the scenario of interest comprises an area of 0.2 $\textrm{km}^2$ in an urban residential area neighborhood. This scenario is characterized by buildings with a height of up to 25 m. These buildings are separated by an avenue with a width varying from 30 to 40 meters, thus forming a street canyon-like environment. In addition, typical obstacles from these environments such as vehicles, vegetation or lampposts are found. Thus, this scenario is prone to a strong multipath channel due to classical propagation phenomena (e.g., scattering and diffraction) along the scenario.

\section{Analysis of the scenario}

\noindent Based on the scenario description, this section discusses radio propagation analysis to study the challenges when FR2 cellular systems are deployed in urban environments.

\subsection{Coverage map and path gain}

As a first approximation, a coverage map of the received power is obtained from the RSRP measured periodically every 20 ms, as set out in the SSB burst set. Fig. \ref{fig1} shows the measured RSRP at several locations with a minimum RSRP level which is 10 dB above the theoretical sensitivity of the R\&S TSME6 radio test scanner. The maximum RSRP is found to be $-92$ dBm in LoS condition for BS-UE distances below 140 m.  This power decreases as: (i)~range increases and (ii)~transition from LoS to NLoS occurs, reaching RSRP values of $-125$ dBm at ranges of 350 m. Note that along the avenue there is a progressive decrease of the signal even in NLoS conditions due to the street canyon effect produced by the buildings. However, when measurements are acquired leaving the main avenue, there is a noticeable drop in the signal level. Within a few meters after turning the corner, the signal drops below the sensitivity of the radio scanner, as denoted by the black lines in Fig. \ref{fig1}.

\begin{figure}[b]
	\centering
	\includegraphics[width= 1\columnwidth]{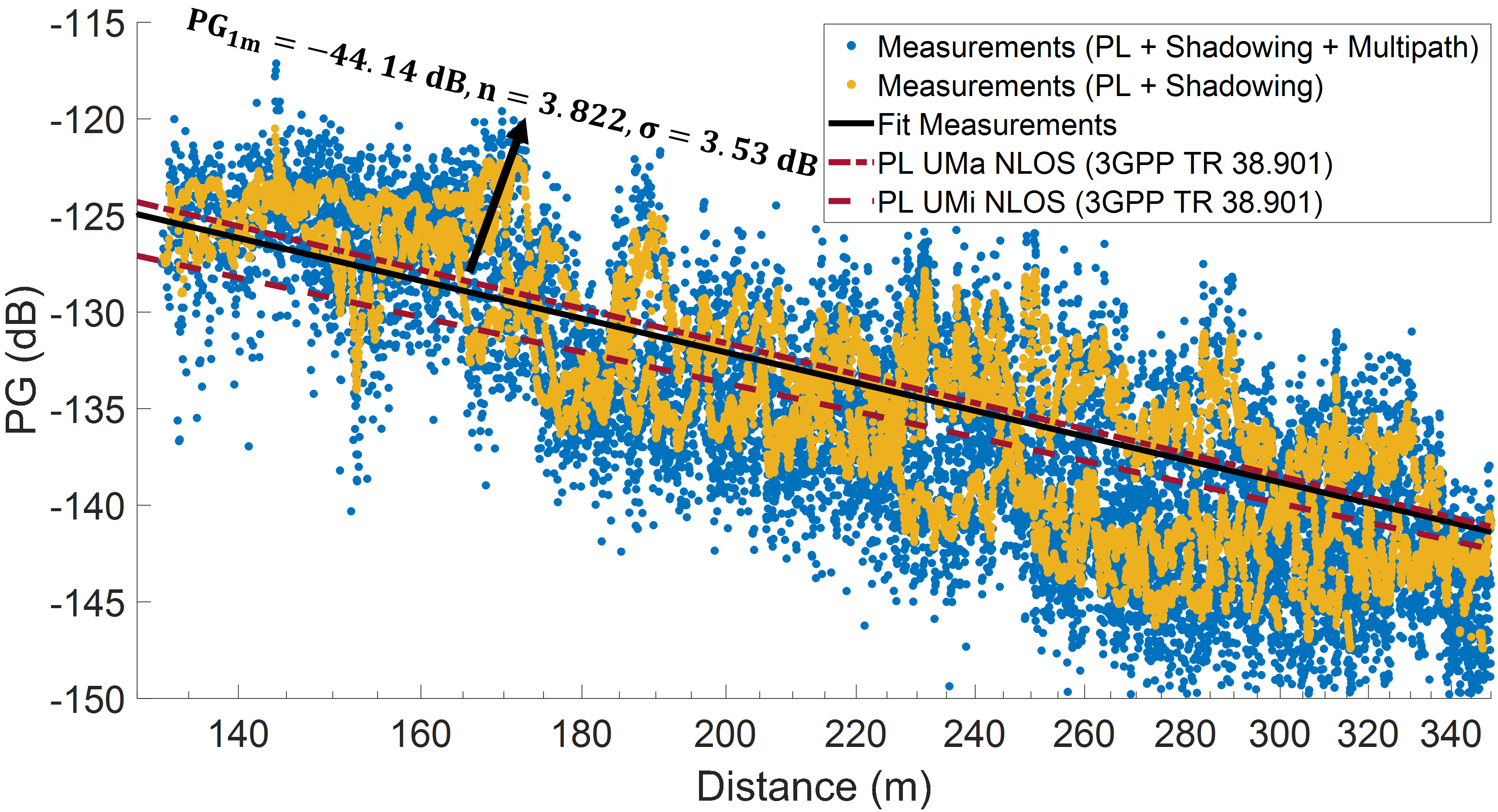}
	\caption{Path gain slope-intercept model given the 3D TX-RX link distance for NLoS case in the main avenue. The yellow dots represent the measurement data after removing the effect of small-scale fading due to multipath.} 
	\label{fig2}
\end{figure}

To characterize the propagation channel, the path gain can be described in terms of the range by following a slope-intercept model as \cite{universal_law}:

\begin{equation}
PG(d)=PG_{1m} - 10 n \log_{10}(d)+\mathcal{N}\left(0, \sigma^2\right),
\end{equation}

\noindent where $PG_{1m}$ is the intercept term, which defines the path gain at 1 m distance, $n$ is the path loss exponent, and $\sigma$ stands for the standard deviation that models a normal distribution representative of the large-scale fading effects in the channel. After compensating the RSRP with the transmitted power in the SSS bandwidth, and the BS and UE antenna nominal gains, the path gain in terms of the distance for NLoS condition is obtained as shown in Fig.~\ref{fig2}. In the FR2 band, where free space losses are already high, a correct estimation of the path gain in NLoS is essential for planning and dimensioning of a deployment. The measured path gain includes the effects due to the distance, large-scale fading, and small-scale fading. To assess the scenario taking into account the effects of large-scale fading, small-scale fading is removed by spatially averaging the measurements over 40$\lambda$ \cite{Lee_theorem}. Thus, the estimated parameters for the slope-intercept model through least-squares optimization are $PG_{1m} = -44.14$ dB, $n = 3.822$, $\sigma = 3.53$~dB. These values match the state-of-the-art models and 3GPP standards that suggest path loss exponents $n \in [3,4]$ for UMa and UMi scenarios with street canyons in the FR2 band \cite{rappaport_UMi, ts38211}. Some of these models are illustrated for comparative purposes in Fig.~\ref{fig2}. Specifically, PG for UMa ($n = 3.91$) and UMi ($n = 3.53$) in NLoS condition given by 3GPP TR38.901 are shown \cite{tr38901}.

\begin{figure}[t]
	\centering
	\includegraphics[width= 1\columnwidth]{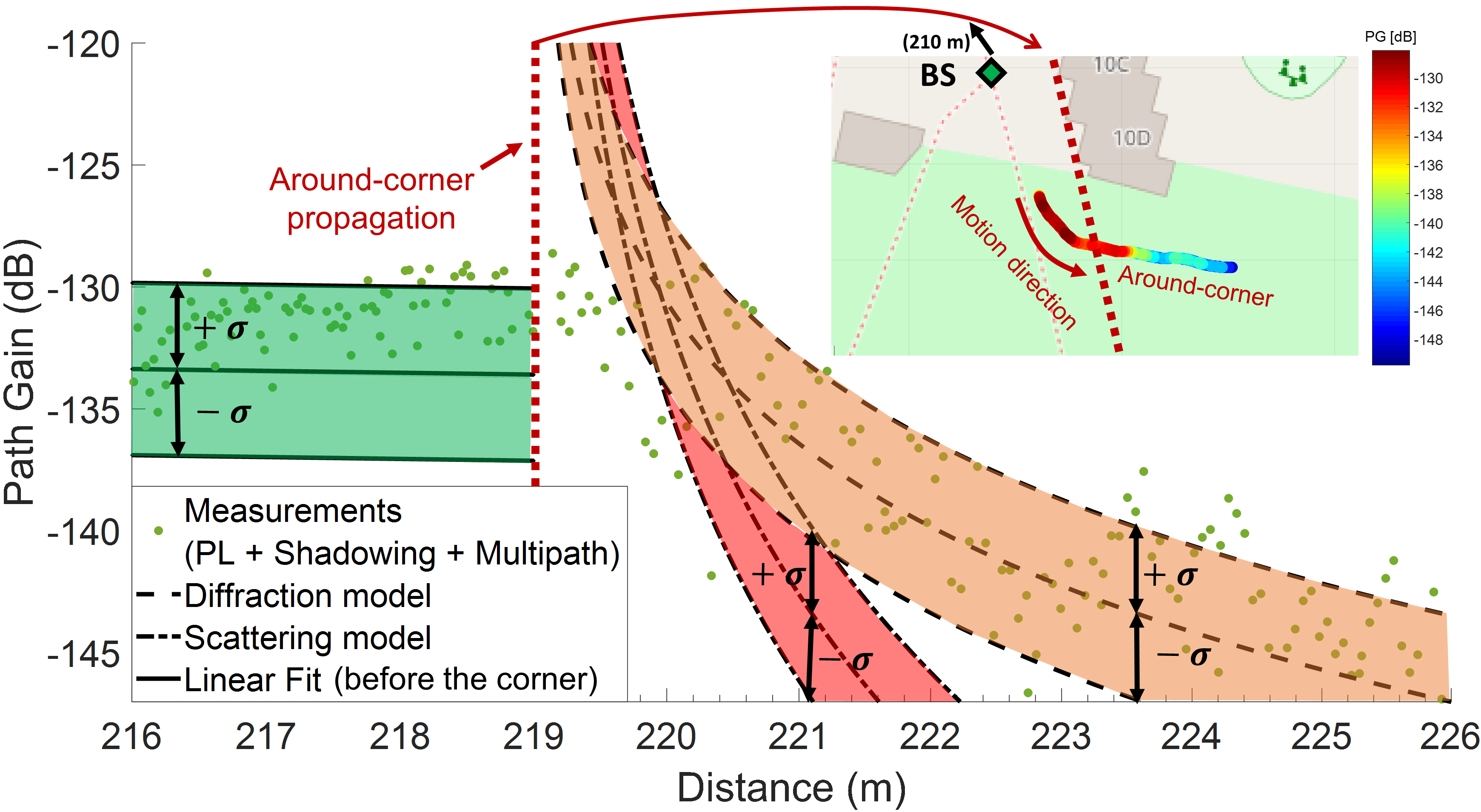}
	\caption{Path Gain for around-corner propagation. Slope-intercept (green) model before the corner, and diffraction (orange) and scattering (red) models when the receiver turns around-corner are illustrated. TX-RX link distance is calculated as the TX-corner plus corner-RX distance, i.e, the Manhattan distance. The direction of the BS with respect to the UE is marked at the top.} 
	\label{fig3}
\end{figure}

\subsection{Around-corner propagation}

Given the previous results, it is reasonable to adjust the path gain in the street canyon through the slope-intercept model due to the low standard deviation. However, around-corner propagation suffers from deep fades that cannot be approximated from a linear fit (see Fig. 1). In these cases, the main around-corner propagation mechanisms can be given by: (i)~scattering, where the corner is assumed to act as a new source, or (ii)~diffraction, where the signal is diffracted at the corner \cite{diffraction_scattering_1}. To determine the predominant propagation mechanism, measurements acquired in around-corner propagation are compared with models from the state-of-the-art for scattering and diffraction phenomena \cite{diffraction_scattering_2, diffraction_scattering_3}. Specifically, Fig. 3 shows a case of around-corner propagation in the micro-cell scenario where the path gain is stable before the UE turn around the corner and suddenly the path gain drops by more than 10 dB after the corner. Thus, scattering and diffraction model curves are presented given $PG_{1m} = -44.14$ dB and $n = 3.822$ values estimated for the slope-intercept model (see Fig. 2) with neglected corner loss~$\Delta$. RMSE is found to be 2.9 dB and 13.1 dB for diffraction and scattering models, respectively. Therefore, the results suggest that diffraction is the main mechanism of around-corner propagation in the scenario. Note the possibility of using the street canyon slope-intercept model to predict the path gain. In summary, street canyon-like environments, such as those found in our scenario, are more likely to experience diffraction as the primary propagation mechanism \cite{diffraction_scattering_3}, attributed to the presence of building structures in open areas. However, according to the state-of-the-art, around-corner propagation in other scenarios may also be predominantly influenced by other propagation mechanisms, such as scattering phenomenon~\cite{diffraction_scattering_2}.

\begin{figure}[t]
	\centering
	\includegraphics[width= 1\columnwidth]{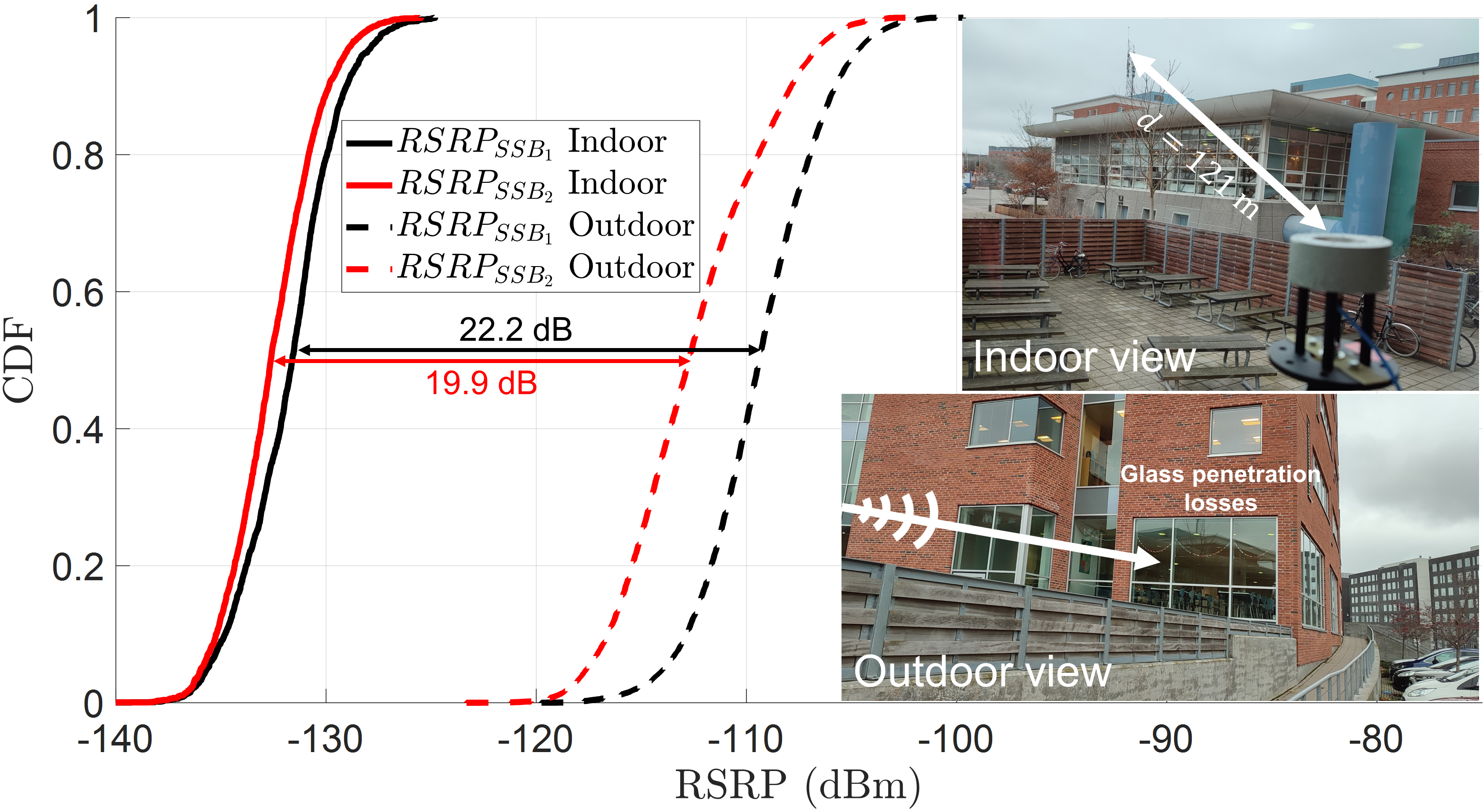}
	\caption{RSRP cumulative distribution function for measurements acquired before and after propagation through a glass in an O2I scenario. Photographs from the outdoor and indoor views are depicted.} 
	\label{fig4}
\end{figure}

\begin{figure}[!b]
	\centering
	\subfigure[]{\includegraphics[width=1\columnwidth]{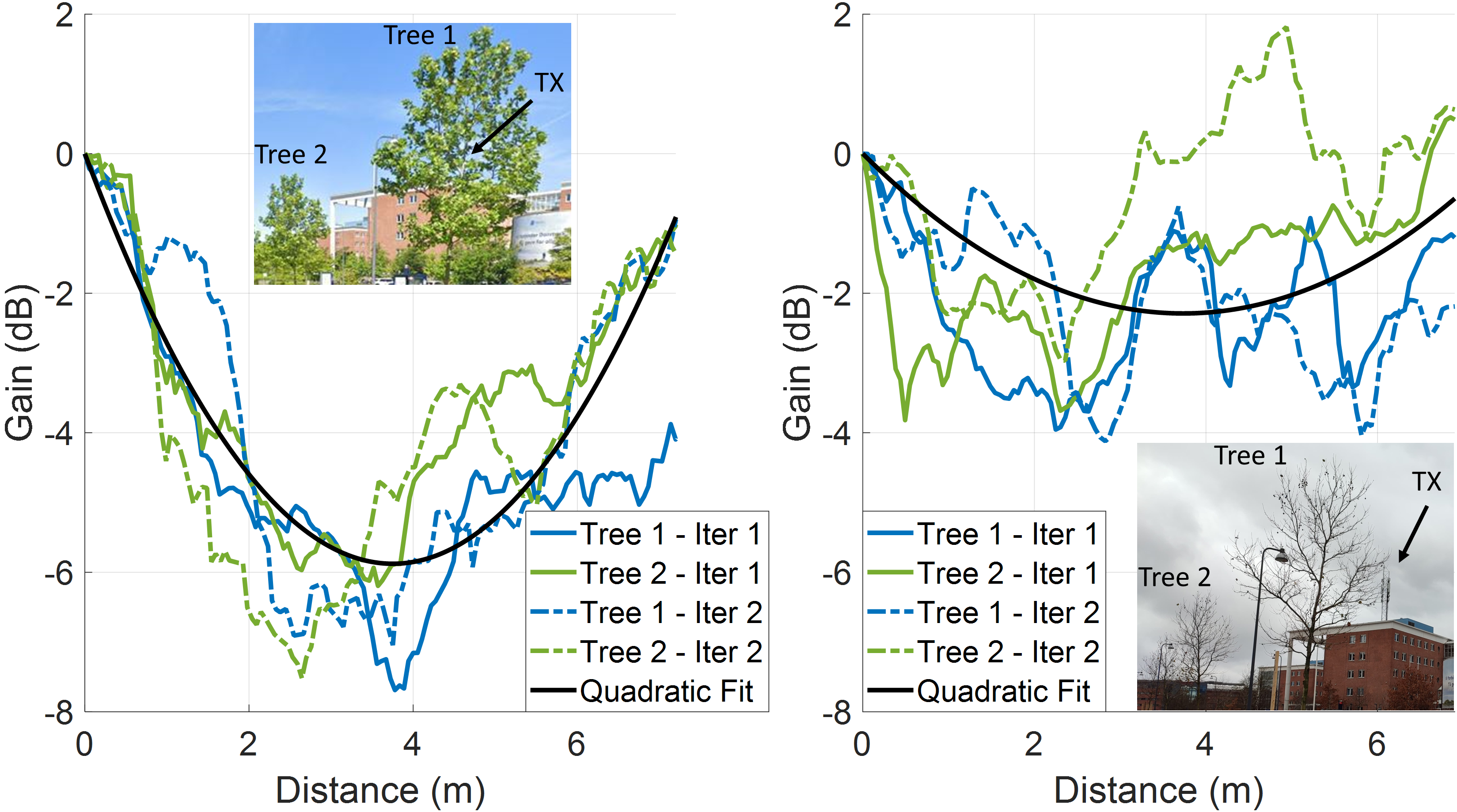}
	}
	\subfigure[]{\includegraphics[width= 1\columnwidth]{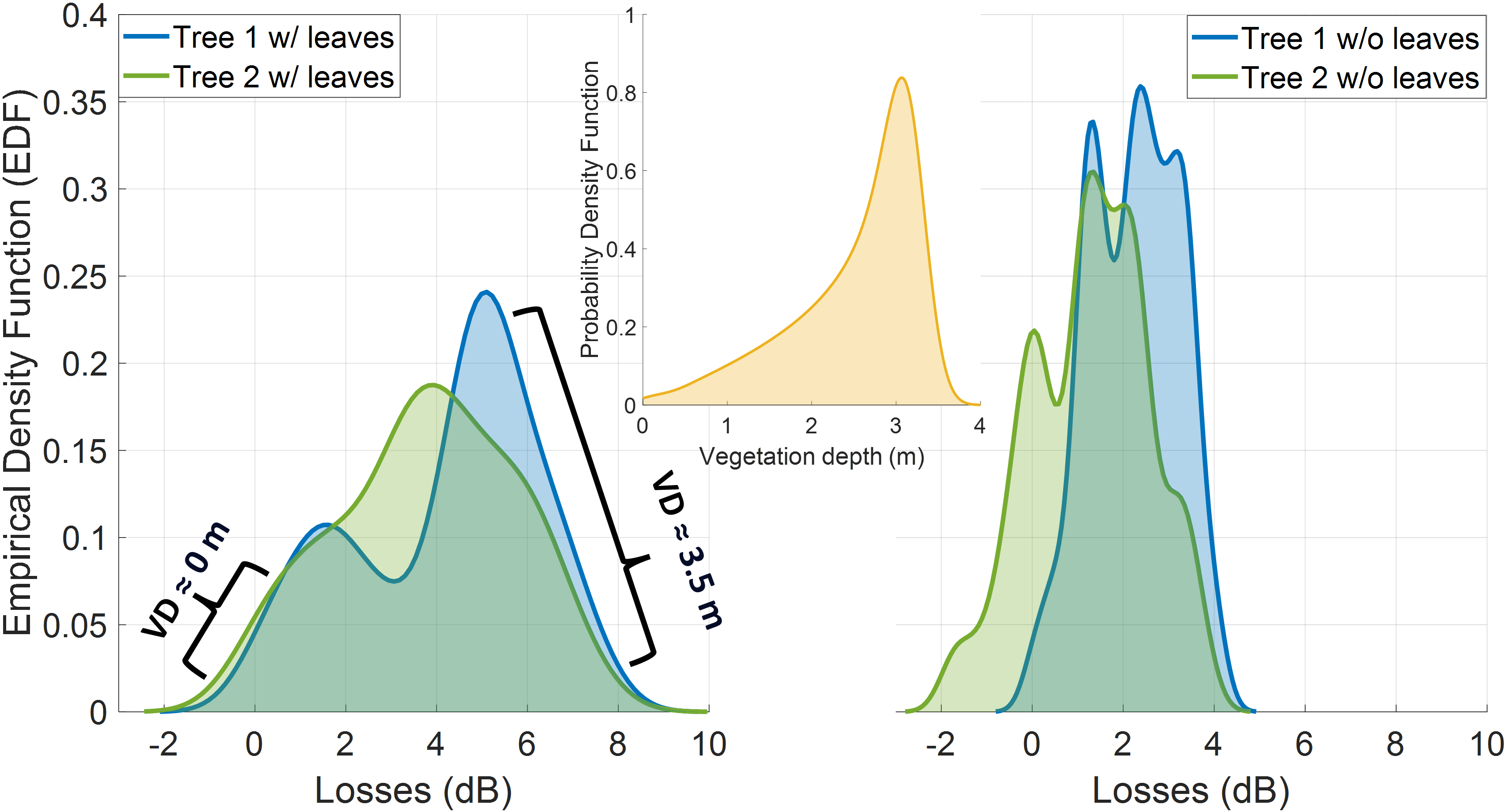}
	}
        \caption{(a) Normalized gain given the vegetation blockage for measurement campaigns performed in October and December. X-axis denotes the path of the UE while the vegetation blocks the LoS. In the case with leaves, the rise and fall periods during blocking are observed. (b) Attenuation losses EDF for both measurement campaigns and vegetation depth (DP) PDF of the analyzed trees.}  
	\label{fig5}
\end{figure}

\subsection{Building penetration losses}

Another issue in urban outdoor scenarios is the possibility of offering in-building coverage through Outdoor-to-Indoor (O2I) propagation. While this is generally possible in the FR1 band, penetration losses in the FR2 band are higher. Therefore, it is essential to assess penetration losses in a live experimental network deployment in the FR2 band. For this purpose, the position in LoS marked in Fig. \ref{fig1} is chosen, where the BS boresight beam is normally incident on the UE. At this location, the RSRP is measured outside and inside the building, as shown in Fig.~\ref{fig4}. Both measurements are separated by a picture window whose size is 4 m $\times$ 2.5 m. This consists of clear double-glazed windows made of an inner and an outer glass layers separated by an air gap. The outer layer, air gap, and inner layer are 8 mm, 16 mm, and 6 mm wide respectively, for a total thickness of 30 mm. A minimum number of 1960 samples is captured for each CDF curve, obtained from several sweeps of the UE along the window. Two SSBs are acquired at different locations within the picture window. The RSRP CDF indicates that a median penetration loss of around \mbox{20-22}~dB is suffered due to the presence of the glass. According to 3GPP TR38.901 \cite{tr38901}, O2I building penetration low-loss and high-loss models at 26~GHz provide penetration loss values of 17.4~dB and 37.3~dB, respectively. Hence, the measurements are in accordance with the standard, tending towards the low-loss model. In the literature there is a large variability in penetration losses due to different types of glass, ranging from values of less than 10~dB, for clear glasses, to 40~dB, for low-emissivity and infrared rejection glasses \cite{glass_1, glass_2, glass_3}. Note that within the micro-cell range, the selected location was the only one where the signal could be reliably measured when the UE was in an indoor location due to its proximity to the BS and the LoS condition. However, the median indoor signal levels are around $-132$ dBm, which is the aggregate effect of signal propagation plus glass attenuation losses. Considering that the sensitivity of our radio test scanner is better than that of conventional UEs \cite{sensitivity_UE}, these results highlight the challenge and difficulty of providing service at these frequencies given the outdoor-to-indoor propagation.

\subsection{Foliage blockage}

As mentioned in Sect. I, not only building penetration losses are relevant in these scenarios, but also street furniture can have a noticeable effect on the received signal. While studying the data in an open space under LoS conditions, anomalous power losses were found. By analyzing the GPS position of the measurements, it was found that this decrease corresponded to the presence of two trees in the LoS link with 3-4 m foliage depth. Note that this measurement campaign was carried out in October, while the leaves of the trees had not fallen yet. To quantify this effect, the measurements were repeated at the same locations in December, when the leaves had already fallen from the trees. Fig. 5(a) shows the normalized gain in the TX-RX link given the tree effect. Images of both scenarios are also shown. A quadratic fit is carried out based on two iterations of the experiment for each of the trees. The results show an average maximum loss of 5.9 dB when leaves are still on the tree, while the average maximum loss decreases to 2.3~dB when seasonal leaf fall occurs. In terms of vegetation depth, Fig. 5(b) presents the estimated probability density function (PDF) for the depth vegetation when the tree obstructs the LoS, and the empirical density functions (EDFs) of the attenuation losses for each tree in the months of October and December. The distributions are consistent with Fig. 5(a) with peaks in the EDF between 4~dB and 6~dB when the trees have leaves, while the distribution is narrower and closer to zero in winter due to seasonal leaf fall. Note that the EDF of attenuation losses with leaves shows a similar trend to the PDF of vegetation depth, indicating the correlation between both parameters. These results suggest the importance of taking vegetation into account at mmWave frequencies. A remarkable dependence on the time of year when deciduous trees are present is also found. Studies in the literature generally report densely vegetated areas with foliage losses above 20 dB \cite{foliage_1,foliage_2}. However, even the presence of individual trees with low leaf density has a remarkable effect, as observed empirically.

\begin{figure}[t]
	\centering
	\includegraphics[width= 1\columnwidth]{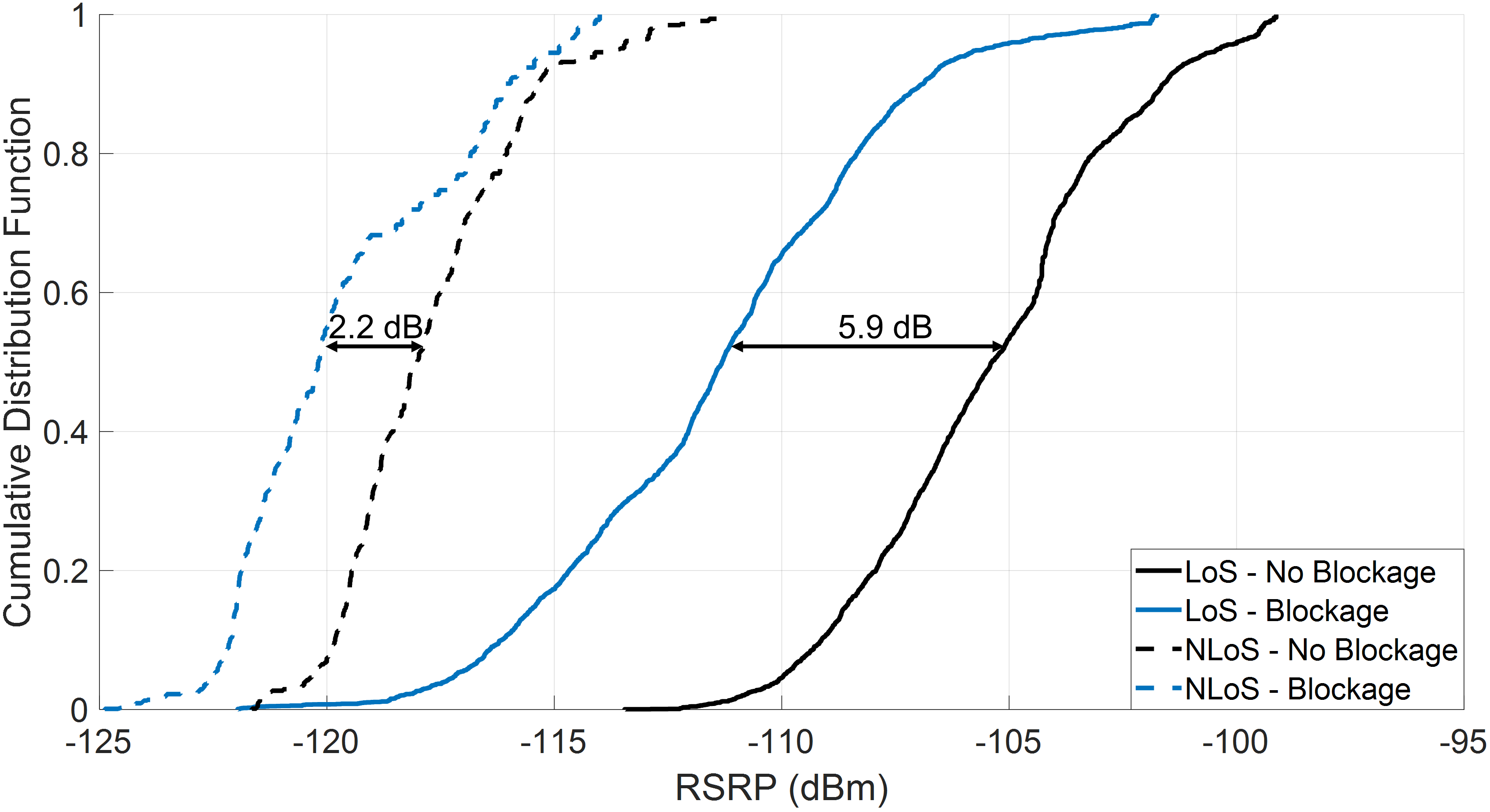}
	\caption{RSRP cumulative distribution function in LoS and NLoS condition with the presence of human blockage losses.} 
	\label{fig6}
\end{figure}

\subsection{Human blockage}

Finally, apart from penetration or vegetation losses, the human body itself can give rise to blockage losses depending on the orientation of the body. Since a subject direction can be any in the outdoor scenario, there are almost certainly times when the person obstructs the communication link. To assess the effect of this blockage, RSRP is measured at multiple locations in the scenario under both LoS and NLoS conditions. At each position, two measurements are acquired: (i) without influence from the human body and (ii) by obstructing the TX-RX link with a human subject. Fig. \ref{fig6} shows both cases for LoS and NLoS. In the LoS case, the median degradation of the RSRP is 5.9 dB, which represents the power losses due to LoS blockage. In the NLoS, the median power losses are 2.2 dB, suggesting that in the absence of a predominant LoS beam, the received power is not reduced as much. These results contrast with those presented in the literature \cite{blockage_1}, where generally higher losses are observed. However, this may be due to the use of directive antennas on short-range communications. When the scenario is based on long-range communications and omnidirectional antennas are used in the UE, the impact of human blockage tends to diminish due to the multipath components in the environment.
 
\section{Conclusions}

\noindent In this letter, a measurement campaign performed in an FR2 live expeimental network deployed in an urban residential area is analyzed.
The study of the campaign outlines some of the key issues in the deployment of such a network. A slope-intercept fit of the data suggests a scenario with high losses due to a path loss exponent $n = 3.822$ in NLoS condition. A sharp drop in power strength is observed when the UE exits the street canyon, with diffraction being the predominant propagation mechanism in perpendicular streets. Additionally, Outdoor-to-Indoor propagation has been examined, revealing penetration losses of approximately 20 dB. Likewise, vegetation-induced losses are 5.9 dB and 2.3 dB for individual trees with and without leaves, respectively. Finally, it is observed that the effect of human blockage is not as pronounced as predicted in the literature, owing to the strong presence of multipath components in long-range communications. 

These observations have been carried out on a FR2 5G live network under realistic operational conditions through a setup that accurately reflects realistic conditions. Hence, valuable insights into the network performance experienced by pedestrians in urban micro-cell environments are provided. Thus, these results are expected to be to aid in the deployment of FR2 mobile networks in urban scenarios.


\begin{thebibliography}{23}

\bibitem{m2150}
\textit{Detailed specifications of the terrestrial radio interfaces of International Mobile Telecommunications-2020 (IMT-2020)}, ITU-R M.2150-1 Recommendation, 2022.

\bibitem{ts38211}
\textit{NR; Physical channels and modulation (Release 17)}, 3GPP Standard TS38.211 V17.4.0, 2023.

\bibitem{ts38213}
\textit{NR; Physical layer procedures for control (Release 17)}, 3GPP Standard TS38.213 V17.4.0, 2023.

\bibitem{channel_modeling_theoretical}
S. Wu, C. -X. Wang, e. -H. M. Aggoune, M. M. Alwakeel and X. You, ``A General 3-D Non-Stationary 5G Wireless Channel Model,'' \textit{IEEE Transactions on Communications}, vol. 66, no. 7, pp. 3065-3078, 2018.

\bibitem{channel_modeling_empirical}
C. -X. Wang, J. Bian, J. Sun, W. Zhang and M. Zhang, ``A Survey of 5G Channel Measurements and Models,'' \textit{IEEE Communications Surveys \& Tutorials}, vol. 20, no. 4, pp. 3142-3168, 2018.

\bibitem{5G_it_will_work}
T. S. Rappaport et al., ``Millimeter Wave Mobile Communications for 5G Cellular: It Will Work!,'' \textit{IEEE Access}, vol. 1, pp. 335-349, 2013.

\vfill\break

\bibitem{diffraction_scattering_3}
J. Du \textit{et al.}, ``Directional Measurements in Urban Street Canyons From Macro Rooftop Sites at 28 GHz for 90\% Outdoor Coverage,'' \textit{IEEE Transactions on Antennas and Propagation}, vol. 69, no. 6, pp. 3459-3469, 2021.

\bibitem{glass_2}
J. Du, D. Chizhik, R. Feick, G. Castro, M. Rodríguez and R. A. Valenzuela, ``Suburban Residential Building Penetration Loss at 28 GHz for Fixed Wireless Access,'' \textit{IEEE Wireless Communications Letters}, vol. 7, no. 6, pp. 890-893, 2018.

\bibitem{foliage_2}
I. Rodriguez, R. Abreu, E. P. L. Almeida, M. Lauridsen, A. Loureiro and P. Mogensen, ``24 GHz cmwave radio propagation through vegetation: Suburban tree clutter attenuation,'' in \textit{2016 10th European Conference on Antennas and Propagation (EuCAP)}, Davos, Switzerland, pp. 1-5, 2016. 

\bibitem{operational_sub6}
L. Chiaraviglio et al., "Massive Measurements of 5G Exposure in a Town: Methodology and Results," \textit{IEEE Open Journal of the Communications Society}, vol. 2, pp. 2029-2048, 2021

\bibitem{ts381014}
\textit{NR; User Equipment (UE) radio transmission and reception; Part 4: Performance requirements (Release 17)}, 3GPP Standard TS38.101-4 V17.7.0, 2023.

\bibitem{TX_antenna}
EMF Test Report: Ericsson AIR 5322 B258/B258A (FCC), Rev. H, 2021-06-03. [Online]. Available: \href{https://fcc.report/FCC-ID/TA8AKRX10103/5304026.pdf}{https://fcc.report/FCC-ID/TA8AKRX10103/5304026.pdf}

\bibitem{universal_law}
D. Chizhik, J. Du and R. A. Valenzuela, ``Universal Path Gain Laws for Common Wireless Communication Environments,'' \textit{IEEE Transactions on Antennas and Propagation}, vol. 70, no. 4, pp. 2928-2941, 2022.

\bibitem{Lee_theorem}
W. C. Y. Lee, ``Estimate of local average power of a mobile radio signal,'' \textit{IEEE Transactions on Vehicular Technology}, vol. 34, no. 1, pp. 22-27, 1985.

\bibitem{rappaport_UMi}
S. Sun \textit{et al.}, ``Propagation Path Loss Models for 5G Urban Micro- and Macro-Cellular Scenarios'' in \textit{2016 IEEE 83rd Vehicular Technology Conference (VTC Spring)}, Nanjing, China, pp. 1-6, 2016.

\bibitem{diffraction_scattering_1}
A. Karttunen, A. F. Molisch, S. Hur, J. Park and C. J. Zhang, ``Spatially Consistent Street-by-Street Path Loss Model for 28-GHz Channels in Micro Cell Urban Environments,'' \textit{IEEE Transactions on Wireless Communications}, vol. 16, no. 11, pp. 7538-7550, 2017.

\bibitem{diffraction_scattering_2}
D. Chizhik, J. Du, R. Feick, M. Rodriguez, G. Castro and R. A. Valenzuela, ``Path Loss and Directional Gain Measurements at 28 GHz for Non-Line-of-Sight Coverage of Indoors With Corridors,'' \textit{IEEE Transactions on Antennas and Propagation}, vol. 68, no. 6, pp. 4820-4830, 2020.

\bibitem{tr38901}
\textit{Study on channel model for frequencies from 0.5 to 100 GHz (Release 17)}, 3GPP Standard TR38.901 V17.0.0, 2022.

\bibitem{glass_1}
H. Zhao \textit{et al.}, ``28 GHz millimeter wave cellular communication measurements for reflection and penetration loss in and around buildings in New York city,'' in \textit{2013 IEEE International Conference on Communications (ICC)}, Budapest, Hungary, pp. 5163-5167, 2013. 

\bibitem{glass_3}
Z. Zhong, J. Zhao and C. Li, ``Outdoor-to-Indoor Channel Measurement and Coverage Analysis for 5G Typical Spectrums'', \textit{International Journal of Antennas and Propagation}, vol. 2019, pp. 3981678, 2019.

\bibitem{sensitivity_UE}
M. López, T. B. Sørensen, I. Z. Kovács, J. Wigard and P. Mogensen, ``Measurement-Based Outage Probability Estimation for Mission-Critical Services," \textit{IEEE Access}, vol. 9, pp. 169395-169408, 2021.

\bibitem{foliage_1}
S. Perras and L. Bouchard, ``Fading characteristics of RF signals due to foliage in frequency bands from 2 to 60 GHz,'' in \textit{The 5th International Symposium on Wireless Personal Multimedia Communications}, Honolulu, HI, USA, pp. 267-271, 2002.

\bibitem{blockage_1}
R. Schulpen, L. A. Bronckers, A. B. Smolders and U. Johannsen, ``Impact of Human Blockage on Dynamic Indoor Multipath Channels at 27 GHz,'' \textit{IEEE Transactions on Antennas and Propagation}, vol. 70, no. 9, pp. 8291-8303, 2022.

\end{thebibliography}
\end{document}